\documentclass[aps,prb,twocolumn,showpacs,superscriptaddress]{revtex4-1} 

\usepackage{graphicx}
\usepackage{dcolumn}
\usepackage{bm}
\usepackage{amsmath}
\usepackage{natbib}
\usepackage{lineno}
\usepackage{array}

\begin{document}

\title{Strong reduction of the coercivity by a surface acoustic wave in an out-of-plane magnetized epilayer}

\author{L. Thevenard$^{1}$\email[e-mail: ]{thevenard@insp.jussieu.fr}, I. S. Camara$^{1}$, J.-Y. Prieur$^{1}$, P. Rovillain$^{1}$, A. Lema\^itre$^{2}$, C. Gourdon$^{1}$,  and J.-Y. Duquesne}

\affiliation{
Sorbonne Universit\'es, UPMC Univ Paris 06, CNRS,  Institut des Nanosciences de Paris, 4 place Jussieu,75252 Paris France\\
$^2$  Laboratoire de Photonique et Nanostructures, CNRS, UPR 20, Route de Nozay, Marcoussis, 91460 France}

\date{\today}

\label{sec:Abstract}

\begin{abstract}

Inverse magnetostriction is the effect by which magnetization can be changed upon application of  stress/strain. A strain modulation may be created electrically by exciting  interdigitated transducers to  generate surface acoustic waves (SAWs). Hence SAWs appear as a possible route towards induction-free undulatory magnetic data manipulation. Here we demonstrate experimentally  on an out-of-plane magnetostrictive layer a reduction of the coercive field of up to 60$\%$ by a SAW,  over millimetric distances.  A simple model shows that this spectacular effect can be partly explained by the periodic  lowering of the strain-dependent domain nucleation energy by the SAW. This proof of concept was done on (Ga,Mn)(As,P), a magnetic semiconductor  in which the out-of-plane magnetic anisotropy can be made very weak by epitaxial growth; it should guide material engineering for all-acoustic magnetization switching.

\end{abstract}


\maketitle


The use of electrical voltage - rather than current - is a promising approach to limit heat dissipation and facilitate device size reduction for magnetic data manipulation. Static  changes to the anisotropy by carrier density modification, magnetoelectricity  or magnetoelasticity (mediated by piezoelectricity) have been demonstrated via gating in metals \cite{Schellekens2012,Shiota2011,Shiota2011,Cavill2013} and magnetic semiconductors  \cite{Chiba2008,Cormier2014,Balestriere2011,Bihler2008}.  For instance,  piezoelectric actuators have enabled the field-less driving of domain walls in the highly magnetostrictive FeGa \cite{Cavill2013}, and the static manipulation of magnetization  in (Ga,Mn)As \cite{Bihler2008}.

Using magnetoelasticity with a time-varying strain, such as surface acoustic waves (SAWs), or bulk picosecond acoustic pulses (PAPs) could allow faster switching. Their wave properties  would also offer the possibility of focusing or interference to switch magnetization selectively, a potentially exciting perspective for magnetic data storage applications. SAWs and PAPs  have  been shown to couple to magnetization in  magnetostrictive materials \cite{Zhou2014,Thevenard14,Scherbakov2010,Jager2013}. Relying on this coupling, recent work has predicted\cite{Kovalenko2013,Thevenard2013} and demonstrated experimentally\cite{Prec2016} that acoustic waves can trigger \textit{resonant}  precessional switching. SAWs can moreover assist - \textit{non resonantly}  -  the coherent reversal of small structures\cite{Davis2010,Davis2015}, or in larger samples could trigger the nucleation  or propagation of domain walls \cite{Dean2015a,Thevenard2013}. In that case, the lever is expected to be the dependency of the domain wall energy on the  magnetoelastic constant. So far the strongest reported effect has been in thin FeGa layers, in which a 11 $\%$ reduction of the coercive field by SAW was observed,  as well as early evidence of localized magnetization reversal assisted by  a focussed SAW\cite{Li2014a}. No particular mechanism was identified however. In this work, we   extend this proof of concept to out-of-plane magnetized materials, as well as to an entirely different class of materials: dilute magnetic semiconductors (DMS). We evidence an increased efficiency (over 50 $\%$ reduction of the coercive field),   and show by a simple model that  SAW-assisted nucleation is partly responsible for this effect. 

Archetype amongst DMS, (Ga,Mn)As and its recently developed alloy (Ga,Mn)(As,P) have a carrier-mediated ferromagnetic phase which renders their magnetic anisotropy sensitive to both static and dynamic strain  \cite{lemaitre08,Dietl2014,Bihler2008}.  Rayleigh waves are   particularly well suited to the magnetoelasticity of  (Ga,Mn)(As,P), since their strain components directly couple to the  uniaxial out-of-plane anisotropy. SAW-induced precessional switching was for instance recently  demonstrated \cite{Prec2016}, under fields of a few hundreds of mT applied perpendicular to the easy axis.  In this work,  we focus  on the geometry where the field is along the easy axis, and  use Kerr microscopy to monitor  modifications to the coercive field and the shape of nucleated domains in the presence of a SAW.

 	\begin{figure*}
	\centering
	\includegraphics[width=0.23\textwidth,angle=90,origin=c]{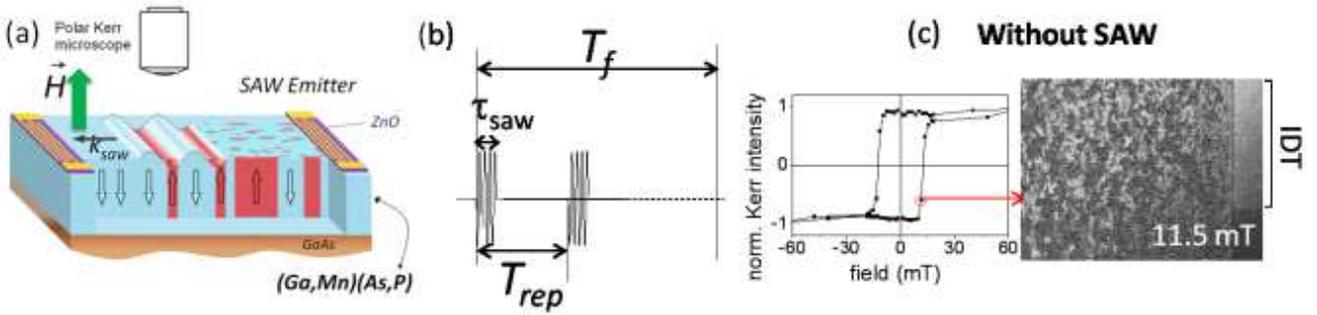}
	\caption{(a) Schematics of the experimental set-up [not to scale]. A thin SiO$_{2}$ layer  [not represented] is deposited on the (Ga,Mn)(As,P). The SAW assists domain nucleation in the half-period during which it \textit{lowers} the DW energy. (b) Time dependence of the strain [not to scale].  SAWs are generated  in bursts of duration $\tau_{SAW}$=600~ns,  every $T_{rep}$=20~ms, during the entire field plateaus of length $T_{f}$=4~s. (c) Hysteresis cycle averaged in front of the emitting IDT without SAW ($T$=10~K) and Kerr microscopy image (365$\times$273 $\mu m^2$) of the domains during the field reversal. The magnetic part of the image has been divided by a reference image taken at remanence and the right-most part left raw to show the transducer.}
	\label{fig:scheme}
\end{figure*}


The sample studied is an out-of-plane magnetized $d$=50~nm thick layer of (Ga$_{1-x}$,Mn$_{x}$)(As$_{0.96}$,P$_{0.04}$) in tensile strain on GaAs with an active Mn concentration $x^{eff}_{Mn}\approx 3.5\%$,  and a  Curie temperature of T$_{C}$ = 95~K.  The SAWs were generated and detected by two opposite sets of gold Interdigitated Transducers (IDTs) evaporated on a SiO$_{2}$/ZnO bilayer, with a 1~mm aperture. The IDT teeth were 1.25~$\mu$m wide and equidistant, giving a SAW wavelength of 5~$\mu$m and frequency of $f_{SAW}$=549~MHz (SAW period $T_{SAW}\approx1.8$~ns).  A window   etched in the ZnO layer between the two IDTs  facilitated the Kerr imaging, with the  SiO$_{2}$ layer thickness adjusted for an optimal Kerr contrast \cite{Riahi2015a} at $\lambda$=600~nm. Whereas the IDTs are separated by 2~mm, the typical field of view for the polar Kerr microscope is between 360 and 994 $\mu$m wide, depending on the magnification used (Fig. \ref{fig:scheme}a).  To limit device break-down at high powers, the SAW is generated in bursts (typically $\tau_{SAW}$=600~ns long) emitted at a low repetition frequency (typically $f_{rep}$=50 Hz). The maximum surface strain has been estimated using a Vector Network Analyzer (VNA) to be of 6.2$\times 10^{-4}$. The magnetic field is applied in plateaus of a few seconds to allow image acquisition and transfer (see Fig. \ref{fig:scheme}b for the time-dependence of the strain). Spatially averaging the Kerr intensity at each field gives the  hysteresis cycle, from which we define the coercive field $B_{c}$ as the one  giving zero average magnetization.

\begin{figure*}
	\centering
	\includegraphics[width=0.3\textwidth,angle=90,origin=c]{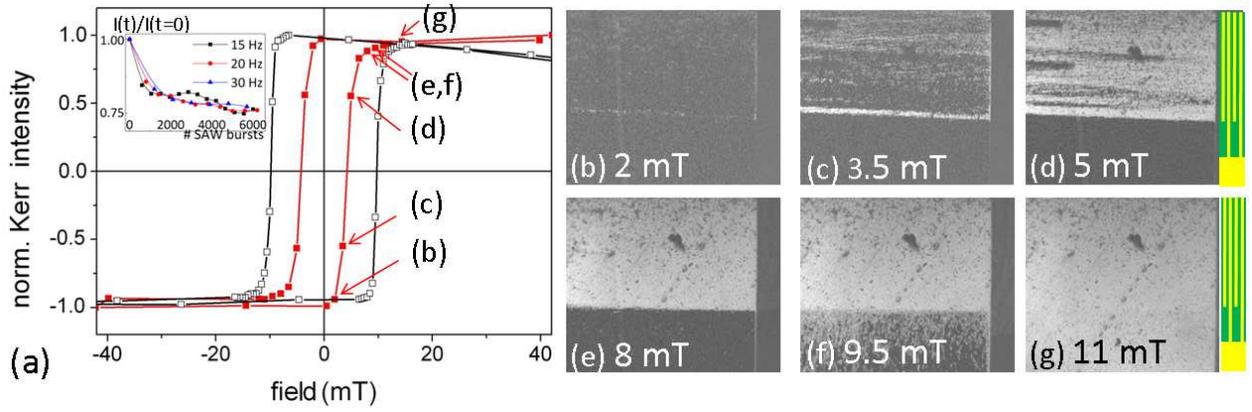}
	\caption{SAW assisted reversal at P$_0$=8.9~W/mm ($T$=30K): (a) Hysteresis cycles averaged in front of the SAW emitter, with and without SAW. Inset: magnetic after-effect experiment at $B$=2~mT under 6000 SAW bursts, corresponding to a duration of 6 min ($f_{rep}$=15 Hz), 5 min ($f_{rep}$=20 Hz) or 3.5 min ($f_{rep}$=30 Hz), intensity averaged in front of the IDT  normalized  to its initial value. (b-g) Kerr microscopy images (690$\times$924 $\mu m^2$) corresponding to the cycle under SAW in (a).}
	\label{fig:domaines}
\end{figure*}

Without SAW,  the hysteresis loop averaged in front of the emitting IDT is square (Fig. \ref{fig:scheme}c), and the  reversal is driven by multiple nucleations across the layer. Note that this behavior is very different from high anisotropy (Ga,Mn)As samples \cite{Thevenard2006}, in which nucleation events are rare, especially at low temperatures, and occur at high enough fields to then trigger the fast reversal of the whole layer.  A spatial mapping of the coercive field  does not highlight any particular nucleation spots. Hysteresis cycles are then taken with the SAW 'on'. Fig. \ref{fig:domaines}a shows for instance data taken at 30K when exciting the IDT at its resonance frequency 549~MHz with an incident rf power P$_0$=8.9~W/mm, and a repetition rate of $f_{rep}$=50 Hz. The coercive field in front of the emitting IDT is now 4.3~mT, reduced by over 50$\%$ compared to the one without SAW  (9.8~mT). The Kerr images  show that switching clearly occurs first on the SAW path (Fig. \ref{fig:domaines}b-e). More specifically it initiates at low field ($<$3.5~mT),  on the   edge of the wave front (Fig. \ref{fig:domaines}b) and with sparse nucleation centres above. At 8~mT, the layer has almost fully switched on the SAW path, but not at all out of it. As the field is further increased to 9.5~mT (Fig. \ref{fig:domaines}f), switching is finally triggered out of the SAW path and the magnetization fully reverses at 11~mT.

 In order to have a first hint as to the reversal mechanism, a magnetic  after-effect experiment is performed, i.e. a study of the time dependence of the magnetic susceptibility after a given initialization. After saturating the layer 'up', a field of opposite sign is applied, low enough to allow slow magnetization reversal ($B$=2~mT, T=30K). Three experiments are then run  under different SAW repetition frequencies ($f_{rep}$=15, 20 and 30 Hz). For each $f_{rep}$, the rf power is applied long enough to total the passage of   6000 SAW bursts. The intensity is averaged in front of the IDT, normalized to its initial value, $I(t)/I(t=0)$, and plotted versus  the number of SAW bursts (inset of Fig. \ref{fig:domaines}a). The  three curves have a very similar shape and final fractional reversed magnetization (75$\%$) regardless of the repetition frequency. This result suggests  a cumulative effect necessary for the switching, given by the total number of SAW periods with a weak influence of the repetition rate.

\begin{figure}
	\centering
	\includegraphics[width=0.2\textwidth,angle=90,origin=c]{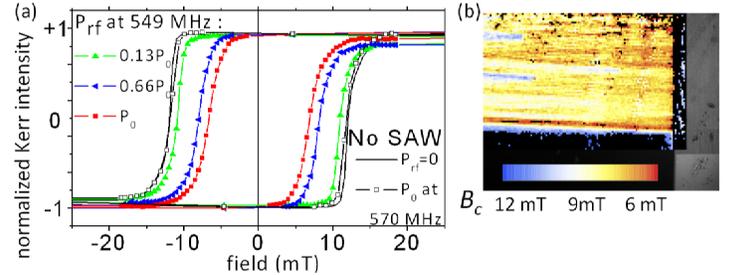}
	\caption{(a) Hysteresis cycles averaged on the SAW path (T=10K), without SAW (full line), off IDT resonance at P$_0$ (open symbols), and at the IDT resonance for different powers.  (b) Spatial map of the coercive field based on the hysteresis cycle taken at P$_0$ (image size 690$\times$924 $\mu m^2$).  Blue (red) indicates a poor (good) SAW efficiency. Transparent bins indicate that the data was locally too poor to extract a coercive field.}
	\label{fig:PandMAP}
\end{figure}

\begin{figure}
	\centering
	\includegraphics[width=0.22\textwidth,angle=90,origin=c]{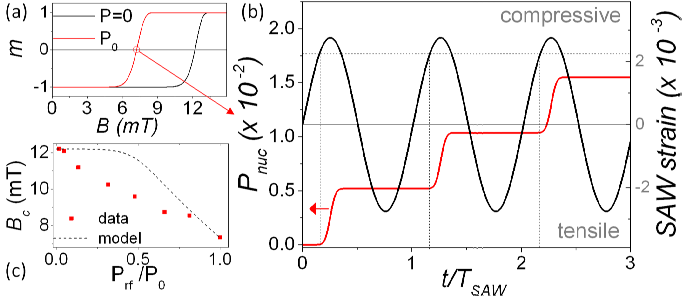}
	\caption{(a) Fraction of reversed magnetization versus field calculated by the model, with and without SAW. (b) Time dependence of the SAW-induced strain and nucleation probability at $B$=7.35 mT. Dotted lines materializes the onset of nucleation triggered by reaching a threshold compressive strain. (c) Power dependence of the coercive field at T=10~K, nucleation model and data, taken as the central value of the Gaussian distribution of $B_c$ in a small area in front of the IDT.}
	\label{fig:modele}
\end{figure}

The excitation power was then varied over two decades (Fig. \ref{fig:PandMAP}a, T=10K). The SAW efficiency clearly decreases  with decreasing power, with the coercive field recovering its SAW-less value at 0.01P$_0$ (cycle not shown). The final $B_{c}$(P) curve is roughly linear in power (Fig. \ref{fig:modele}a), and exciting SAWs with the left IDT gives an identical trend.  No $B_{c}$ reduction is observed when  exciting the IDT off resonance, for which no SAW is generated (rf power at 570~MHz, open symbols in Fig. \ref{fig:PandMAP}a), proving the effect is not due to the  field radiated by the antenna. To obtain the spatial dependence of the coercive field, the Kerr intensity is averaged over 10x10 pix$^2$ (6.6$\times$6.6 $\mu$m$^{2}$) bins for each field. The resulting bin-specific hysteresis cycles are then analysed numerically to extract a local coercive field which is overlaid on the original image as a color map, where blue (red) indicates a poor (good) SAW efficiency (Fig. \ref{fig:PandMAP}b). The SAW efficiency is clearly not uniform, because of both  a complex wave-front shape due to the finite aperture of the IDT, and  defects on the sample. Large defects on the surface of the sample  "shadow" the SAW (upper left-hand corner), giving a  higher (bluer) coercive field in their wake. The resulting spatial dispersion of the  SAW efficiency also explains the broadening of the switching step with rf power seen in  the hysteresis cycles averaged over the whole  SAW path  (Fig. \ref{fig:PandMAP}a). 

\begin{figure}
	\centering
	\includegraphics[width=0.31\textwidth,angle=90,origin=c]{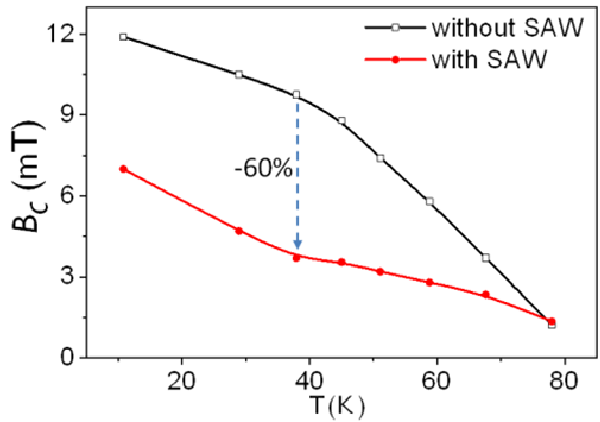}
	\caption{Temperature dependence of the coercive field averaged  in front of the IDT (incident rf power P$_0$).}
	\label{fig:T}
\end{figure}

\begin{figure}
	\centering
	\includegraphics[width=0.15\textwidth,angle=90,origin=c]{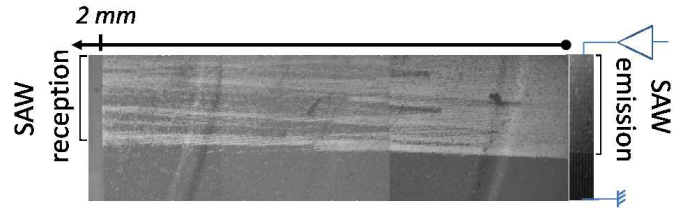}
	\caption{Kerr microscopy images (690$\times$924 $\mu m^2$) taken at P$_0$, T=10 K and  $B$=11~mT. The vertical streaks in the first and last images are microscopy artefacts.}
	\label{fig:att}
\end{figure}

The efficiency of this coercivity reduction was then studied as a function of temperature. Hysteresis cycles were taken   up to 80~K (15 K below T$_{C}$), beyond which the Kerr contrast was too poor for proper analysis (Fig. \ref{fig:T}). The efficiency of the SAW-assisted reversal  is roughly constant up to 38 K (5-6 mT reduction of $B_c$)  and then decreases, before cancelling out at about 70~K. This may be due to a concomitant decrease of the IDT's electro-mechanical conversion efficiency. This was indeed observed using a  VNA.  Finally, we  studied the distance up to which the SAW assists magnetization reversal (T=~10K). Three hysteresis cycles were taken,  moving for each one the sample in front of the objective so as to observe the whole 2~mm long delay line between emitting and receiving transducers whilst the SAW was 'on'. The resulting stitched panorama shows that the SAW stays efficient over millimetric distances (Fig. \ref{fig:att}).

We now review the possible mechanisms responsible for
the spectacular effect of the surface acoustic wave on the
coercivity. We first discuss why thermal effects can be discarded.
Indeed, rf excitation of an IDT can possibly generate heat because of Joule
effect in the transducer or Foucault electrical losses in the metallic bodies submitted to the radiated rf electromagnetic fields. The spectrum of thermal excitation will comprise a DC component (due to the mean input power) and high frequency components in the GHz range (twice the  rf frequency). The DC component will be isotropic and therefore cannot account for the observed highly directive magnetization reversal. Moreover, the DC component depends on the repetition rate so that it can be monitored: measuring the acoustic velocity change versus repetition rate,
the temperature rise was estimated to be under 0.5~K (for a base temperature of 30~K). High frequency components will exibit very short diffusion lengths, in the $\mu$m range depending on diffusivities, and can therefore not account for the observed effect over millimetric distances. The acoustic wave could also induce a temperature rise while propagating because part of the acoustic loss results in incoherent phonons production (thermoelastic, Akhieser effects\cite{Maris1971} etc.).
This effect is in fact usually negligeable. In our case, crudely assuming that the entire lost acoustic energy is dissipated as heat, we estimate that the temperature rise due to those processes lies below 0.05~K.

We therefore focus on non-thermal effects of the SAW. With the field  along the easy axis, magnetization reversal proceeds by the nucleation and propagation of domain walls (DWs). With $A_{ex}$   the exchange constant\cite{Shihab2015} and $K_{u}$ the out-of-plane uniaxial anisotropy (measured by cavity FMR), their energy is given by $\sigma$=$4\sqrt{A_{ex}K_u}$, and their velocity in the stationary regime \cite{malo79} by $v_{stat}\propto \Delta$, with $\Delta$=$\sqrt{A_{ex}/K_u}$. Both $\sigma$ and $v_{stat}$ depend on $K_{u}$,   the dominant magnetoelastic term in (Ga,Mn)(As,P). The SAW could thus - in theory - affect both steps of the reversal.  It has been argued however\cite{Davis2015} that the high frequencies of SAWs ($>$100~MHz) would not allow enough time for the relatively slow  DW propagation  (speeds of tens of ms$^{-1}$ in GaMnAs \cite{Dourlat2008}) - unless particular geometries are used\cite{Dean2015a} - so we focus on their effect on  nucleation.   The concavity of the magnetic after-effect curves taken under SAW (inset of Fig. \ref{fig:domaines}a.) supports a nucleation-dominated reversal\cite{Labrune1989}. We therefore  model the influence of the SAW on  domain nucleation, and how it may lead to the  power dependence of the coercivity reduction observed in Fig. \ref{fig:modele}a.

Following Hubert \textit{et al.} \cite{Hubert},  the nucleation of an isolated domain of radius $r_{nuc} $ is driven by the competition between the  cost to create a DW, $ \Delta E_{nuc}$=$2\pi r_{nuc}d\sigma $, and the associated lowering of the Zeeman energy\footnote{In very soft materials where magnetization reverses \textit{before} a magnetic field sign change, a demagnetizing contribution must also be included \cite{Hubert} - it will be neglected here given the shape of the hysteresis loop (Fig. \ref{fig:scheme}c)} $\Delta E_{Z}$=$-2M_{s}Bd\pi r_{nuc}^{2}$, with $ \Delta E_{tot} $=$\Delta E_{Z}+\Delta E_{nuc}$. We remind that $d$ is the layer thickness. This nucleation is thermally activated, with typical switching times given by $\tau=\tau_{0}exp[\frac{\Delta E_{tot}}{k_{B}T}]$, $1/\tau_0$ being an attempt frequency\cite{Brown1963} ($\tau_0 \approx$ 10 ps). The so-called "droplet model" \cite{Barbara1994} argues that the nucleation radius is well approximated by the one leading to  $\frac{\partial\Delta E_{tot}}{\partial r_{nuc}}=0$, i.e.  $r_{nuc}=\frac{\sigma}{2BM_{s}}$. This allows to rewrite the  total energy barrier as: $ \Delta E_{tot}$=$\frac{\pi d\sigma^{2}}{2BM_{s}}$. In our particular case, the DW energy  is time-dependent, with $\sigma(t)$=$4\sqrt{A_{ex}K_u(t)}$. $K_{u}(t)$=-$\frac{M_sA_{2\varepsilon}}{2}[\varepsilon_{0}+\varepsilon(t)]$  is a function of the static and  dynamic (SAW induced) strains\footnote{Since we are in a propagating and not stationary geometry for the SAWs, we will drop the spatial dependence of the dynamic strain.}, with $\varepsilon_{0}$=$\varepsilon_{zz,0}-\varepsilon_{xx,0}$=-0.161$\%$, $A_{2\varepsilon}$=34~T\cite{Thevenard14}  and $\varepsilon(t)$=$\varepsilon_{ZZ}(t)-\frac{\varepsilon_{XX}(t)}{2}$=$\varepsilon_{SAW} \sin\omega_{SAW}t$. The DW nucleation  energy barrier is then  time-dependent  so that the probability to nucleate a domain during a field plateau of duration $T_{f}$ is given by:

 $	P_{nuc}(T_{f},B)=1-exp\left( -\int_0^{T_{f}}exp[-\frac{\Delta E_{tot}(t,B)}{k_{B}T}]\frac{dt}{\tau_0} \right) $.

If the reversal is indeed mainly driven by nucleation, and that these nucleation sites do not interact, the probability $P_{nuc}$=0.5 to nucleate at a given site is then equivalent to reversing half the   total magnetization: $m$=$M(B)/M_s$=0 with $m(B)$= $1-2P_{nuc}(B)$.  Using the sample's 10~K micromagnetic parameters, we  calculate numerically $m(B)$ for each field of the cycle (Fig. \ref{fig:modele}b), and define the coercive field by $m(B_c)$=0. Since the energy barrier is several eV high, much larger than the thermal energy, we assume like others  \cite{Moritz2005a} a localized lowering of the DW energy  $\sigma^{*}$=$\gamma\sigma$ due to defects or magnetic anisotropy inhomogeneities. In the absence of SAW ($P_{rf}$=0, Fig. \ref{fig:modele}a,b)  $B_{c,0}$=12 mT  yields  $\gamma$=5.4$\times$10$^{-3}$. This corresponds to a nucleation radius of $r_{nuc}=\frac{\sigma^{*}}{2BM_{s}}$=5.5 nm, i.e. of the order of the DW width $\Delta\approx$6~nm at 10 K. This value is kept for the calculation in presence of SAW. 

   We then implement numerically the experimental SAW time dependence (Fig. \ref{fig:scheme}b), and take the SAW amplitude as\cite{royer2000elastic}  $\varepsilon_{SAW}$=$\eta\sqrt{P_{rf}}$.  $\eta$ is a measure of the electromechanical conversion efficiency, and of the impedance matching of our system. It is linear with injected rf power up to $P_0$ and is then the only unknown parameter; its value is adjusted so as to give the observed coercive field at a given rf power. For instance running the calculation at $P_0$ requires $\eta$=2.92$\times$10$^{-5}$~(W m$^{-1}$)$^{-1/2}$ to obtain B$_{c}$=7.35 mT. Plotting the time-dependence of $P_{nuc}(t,B_{c})$  shows that nucleation occurs when the SAW crosses a threshold compressive strain (Fig. \ref{fig:modele}b), for which the total out-of-plane anisotropy, and therefore the DW energy, are briefly lowered with respect to their base value. This  facilitates magnetization switching\cite{Thevenard2013} in a nucleation-dominated reversal process, as observed experimentally. With $\eta$=2.92$\times$10$^{-5}$~(W m$^{-1}$)$^{-1/2}$,  the coercive fields at all the other powers are calculated, and the resulting curve is shown in Fig. \ref{fig:modele}c. The model predicts  the coercive field to start decreasing linearly with power for $P_{rf}>0.4P_0$. 
  The value of $\eta$ determined by this approach corresponds to a maximum transient $K_{u}$  lowering of 31$\%$  (at $P_0$). It gives a  SAW amplitude of $\eta\sqrt{P_{0}}$=2.7$\times$10$^{-3}$, reasonably  about 4 times larger than the value determined electrically  with   the  VNA.

   Although the model is a good qualitative demonstration   that the SAW does efficiently assist domain nucleation, it fails  to reproduce the experimentally observed slope.  The impact of the SAW on DW  propagation  can probably not be fully neglected in the analysis. It would act through the  transient increase of the DW width   $\Delta\propto 1/\sqrt{K_{u}(\varepsilon_{SAW})}$, similar to the observations of Shepley \textit{et al.}\cite{Shepley2015} under static strain variations. The probabilistic calculation of SAW-assisted nucleation-propagation  reversal is beyond the scope of this paper, but could for instance be treated in the framework derived by Fatuzzo and Labrune for  ferroelectric domains \cite{Fatuzzo1962,Labrune1989}, once the $v(B)$ curve has been determined.


To summarize, we have demonstrated up to $ 60\% $ coercivity reduction of a thin out-of-plane magnetized (Ga,Mn)(As,P) layer using 549~MHz SAWs.  As shown by a rudimentary model, the most likely mechanism involves SAW-assisted domain nucleation, made possible by the transient reduction of the domain wall energy, although it does not entirely account for the great efficiency of the SAW. It is likely the mechanism at play in the data presented by Li \textit{\textit{\textit{et al.}}}  in  Galfenol \cite{Li2014a}.  What makes this mechanism particularly efficient in (Ga,Mn)(As,P) is the  low  magnetic anisotropy and weak exchange constant of this material, which give a  small domain nucleation energy.  Lower SAW frequencies would probably make it even more efficient:  the SAW amplitude would be higher at equivalent IDT aperture and excitation power, and the layer would have more time per SAW period to nucleate, while profiting from the wave properties of SAWs.

This work was performed in the framework of the SPINSAW project (ANR 13-JS04-0001-01). We acknowledge M. Bernard and S. Majrab for the technical assistance (INSP), J. von Bardeleben (INSP) for the cavity FMR data, and J. Moritz for fruitful discussions.

\bibliographystyle{phjcp}

\begin{thebibliography}{10}

\bibitem{Schellekens2012}
{\sc a.J. Schellekens}, {\sc a.~van~den Brink}, {\sc J.~Franken}, {\sc
  H.~Swagten}, and {\sc B.~Koopmans},
\newblock {\em Nat. Comm.} {\bf 3}, 847 (2012).

\bibitem{Shiota2011}
{\sc Y.~Shiota}, {\sc T.~Nozaki}, {\sc F.~Bonell}, {\sc S.~Murakami}, {\sc
  T.~Shinjo}, and {\sc Y.~Suzuki},
\newblock {\em Nat. Mat.} {\bf 11}, 39 (2011).

\bibitem{Cavill2013}
{\sc S.~Cavill}, {\sc D.~E. Parkes}, {\sc J.~Miguel}, {\sc S.~S. Dhesi}, {\sc
  K.~W. Edmonds}, {\sc R.~P. Campion}, and {\sc A.~W. Rushforth},
\newblock {\em Appl. Phys. Lett.} {\bf 102}, 032405 (2013).

\bibitem{Chiba2008}
{\sc D.~Chiba}, {\sc M.~Sawicki}, {\sc Y.~Nishitani}, {\sc Y.~Nakatani}, {\sc
  F.~Matsukura}, and {\sc H.~Ohno},
\newblock {\em Nature} {\bf 455}, 515 (2008).

\bibitem{Cormier2014}
{\sc M.~Cormier}, {\sc V.~Jeudy}, {\sc T.~Niazi}, {\sc D.~Lucot}, {\sc
  M.~Granada}, {\sc J.~Cibert}, and {\sc A.~Lema{\^{\i}}tre},
\newblock {\em Phys. Rev. B} {\bf 174418}, 1 (2014).

\bibitem{Balestriere2011}
{\sc P.~Balestrière}, {\sc T.~Devolder}, {\sc J.-V. Kim}, {\sc P.~Lecoeur},
  {\sc J.~Wunderlich}, {\sc V.~Novák}, {\sc T.~Jungwirth}, and {\sc
  C.~Chappert},
\newblock {\em Appl. Phys. Lett.} {\bf 99}, 242505 (2011).

\bibitem{Bihler2008}
{\sc C.~Bihler}, {\sc M.~Althammer}, {\sc A.~Brandlmaier}, {\sc
  S.~Gepr{\"{a}}gs}, {\sc M.~Weiler}, {\sc M.~Opel}, {\sc W.~Schoch}, {\sc
  W.~Limmer}, {\sc R.~Gross}, {\sc M.~S. Brandt}, and {\sc S.~T.~B.
  Goennenwein},
\newblock {\em Phys. Rev. B} {\bf
  78}, 45203 (2008).

\bibitem{Zhou2014}
{\sc H.~Zhou}, {\sc a.~Talbi}, {\sc N.~Tiercelin}, and {\sc O.~{Bou Matar}},
\newblock {\em Appl. Phys. Lett.} {\bf 104}, 11 (2014).

\bibitem{Thevenard14}
{\sc L.~Thevenard}, {\sc C.~Gourdon}, {\sc J.~Y. Prieur}, {\sc H.~J. von
  Bardeleben}, {\sc S.~Vincent}, {\sc L.~Becerra}, {\sc L.~Largeau}, and {\sc
  J.-Y. Duquesne},
\newblock {\em Physical Review B} {\bf 90}, 094401 (2014).

\bibitem{Scherbakov2010}
{\sc A.~Scherbakov}, {\sc A.~Salasyuk}, {\sc T.~Akazaki}, {\sc X.~Liu}, {\sc
  M.~Bombeck}, {\sc C.~Br{\"{u}}ggemann}, {\sc D.~Yakovlev}, {\sc V.~Sapega},
  {\sc J.~Furdyna}, and {\sc M.~Bayer},
\newblock {\em Phys. Rev. Lett.} {\bf 105}, 117204 (2010).

\bibitem{Jager2013}
{\sc J.~V. Jäger}, {\sc A.~Scherbakov}, {\sc T.~L. Linnik}, {\sc D.~R.
  Yakovlev}, {\sc M.~Wang}, {\sc P.~Wadley}, {\sc V.~Holy}, {\sc S.~Cavill},
  {\sc T.~Akazaki}, {\sc a.~W. Rushforth}, and {\sc M.~Bayer},
\newblock {\em Appl. Phys. Lett.} {\bf 103}, 032409 (2013).

\bibitem{Kovalenko2013}
{\sc O.~Kovalenko}, {\sc T.~Pezeril}, and {\sc V.~V. Temnov},
\newblock {\em Phys. Rev. Let.} {\bf 110}, 266602 (2013).

\bibitem{Thevenard2013}
{\sc L.~Thevenard}, {\sc J.-Y. Duquesne}, {\sc E.~Peronne}, {\sc H.~J. von
  Bardeleben}, {\sc H.~Jaffr{\`{e}}s}, {\sc S.~Ruttala}, {\sc J.-M. George},
  {\sc A.~Lema{\^{\i}}tre}, and {\sc C.~Gourdon},
\newblock {\em Phys. Rev. B} {\bf 87}, 144402 (2013).

\bibitem{Prec2016}
{\sc {L. Thevenard, I. S. Camara, P. Rovillain, A.
  Lema{\^{i}}tre, C. Gourdon, and J.-Y. Duquesne, under review.}}

\bibitem{Davis2010}
{\sc S.~Davis}, {\sc A.~Baruth}, and {\sc S.~Adenwalla},
\newblock {\em Appl. Phys. Lett.} {\bf 97}, 232507 (2010).

\bibitem{Davis2015}
{\sc S.~Davis}, {\sc J.~a. Borchers}, {\sc B.~B. Maranville}, and {\sc
  S.~Adenwalla},
\newblock {\em J. Appl. Phys.} {\bf 117}, 063904 (2015).

\bibitem{Dean2015a}
{\sc J.~Dean}, {\sc M.~T. Bryan}, {\sc J.~D. Cooper}, {\sc A.~Virbule}, {\sc
  J.~E. Cunningham}, and {\sc T.~J. Hayward},
\newblock {\em Appl. Phys. Lett.} {\bf 107}, 142405 (2015).

\bibitem{Li2014a}
{\sc W.~Li}, {\sc B.~Buford}, {\sc A.~Jander}, and {\sc P.~Dhagat},
\newblock {\em Physica B} {\bf 448}, 151 (2014).


\bibitem{lemaitre08}
{\sc A.~Lema{\^{\i}}tre}, {\sc A.~Miard}, {\sc L.~Travers}, {\sc O.~Mauguin}, {\sc
  L.~Largeau}, {\sc C.~Gourdon}, {\sc V.~Jeudy}, {\sc M.~Tran}, and {\sc
  J.~George},
\newblock {\em Appl. Phys. Lett.} {\bf 93}, 21123 (2008).

\bibitem{Dietl2014}
{\sc T.~Dietl} and {\sc H.~Ohno},
\newblock {\em Rev. Mod. Phys.} {\bf 86}, 187 (2014).

\bibitem{Riahi2015a}
{\sc H.~Riahi}, {\sc L.~Thevenard}, {\sc M.~Maaref}, {\sc B.~Gallas}, {\sc
  A.~Lema{\^{\i}}tre}, and {\sc C.~Gourdon},
\newblock {\em J. Mag.  Magn. Mat.} {\bf 395}, 340
  (2015).

\bibitem{Thevenard2006}
{\sc L.~Thevenard}, {\sc L.~Largeau}, {\sc O.~Mauguin}, {\sc G.~Patriarche},
  {\sc A.~Lema{\^{\i}}tre}, {\sc N.~Vernier}, and {\sc J.~Ferr{\'{e}}},
\newblock {\em Phys. Rev. B} {\bf 73}, 195331 (2006).

\bibitem{Maris1971}
{\sc H.~Maris},
\newblock {\em {Physical Acoustics, edited by W.P. Mason and R. N. Thurston }},
\newblock Academic Press, New York and London, 1971, Vol.VIII, p. 279.

\bibitem{Shihab2015}
{\sc S.~Shihab}, {\sc H.~Riahi}, {\sc L.~Thevenard}, {\sc H.~J. von
  Bardeleben}, {\sc A.~Lema{\^{\i}}tre}, and {\sc C.~Gourdon},
\newblock {\em Appl. Phys. Lett.} {\bf 106}, 142408 (2015).

\bibitem{malo79}
{\sc A.~P. Malozemofff} and {\sc J.~C. Slonczewski},
\newblock {\em {Magnetic domain walls in bubble materials}},
\newblock Academic press, New York, 1979.

\bibitem{Dourlat2008}
{\sc A.~Dourlat}, {\sc V.~Jeudy}, {\sc A.~Lema{\^{\i}}tre}, and {\sc C.~Gourdon},
\newblock {\em Phys. Rev. B} {\bf
  78}, 161303 (2008).

\bibitem{Labrune1989}
{\sc M.~Labrune}, {\sc S.~Andrieu}, {\sc F.~Rio}, and {\sc P.~Bernstein},
\newblock {\em J. Mag.  Magn. Mat.} {\bf 80}, 211
  (1989).

\bibitem{Hubert}
{\sc A.~Hubert} and {\sc R.~Sch{\"{a}}fer},
\newblock {\em {Magnetic domains}},
\newblock Berlin, Springer edition, 2000.

\bibitem{Note1}
In very soft materials where magnetization reverses \protect \textit {before} a
  magnetic field sign change, a demagnetizing contribution must also be
  included \cite {Hubert} - it will be neglected here given the shape of the
  hysteresis loop (Fig. \ref {fig:scheme}c).

\bibitem{Brown1963}
{\sc W.~F. Brown},
\newblock {\em Phys. Rev.} {\bf 130}, 1677 (1963).

\bibitem{Barbara1994}
{\sc B.~Barbara},
\newblock {\em J. Mag.  Magn. Mat.} {\bf 129}, 79
  (1994).

\bibitem{Note2}
Since we are in a propagating and not stationary geometry for the SAWs, we will
  drop the spatial dependence of the dynamic strain.

\bibitem{Moritz2005a}
{\sc J.~Moritz}, {\sc B.~Dieny}, {\sc J.~Nozi{\`{e}}res}, {\sc Y.~Pennec}, {\sc
  J.~Camarero}, and {\sc S.~Pizzini},
\newblock {\em Phys. Rev. B} {\bf 71}, 2 (2005).

\bibitem{royer2000elastic}
{\sc D.~Royer} and {\sc E.~Dieulesaint},
\newblock {\em {Elastic Waves in Solids I: Free and Guided Propagation}},
\newblock Advanced Texts in Physics, Springer, 2000.

\bibitem{Shepley2015}
{\sc P.~M. Shepley}, {\sc A.~W. Rushforth}, {\sc M.~Wang}, {\sc G.~Burnell},
  and {\sc T.~A. Moore},
\newblock {\em Sci. Rep.} {\bf 5}, 7921 (2015).

\bibitem{Fatuzzo1962}
{\sc E.~Fatuzzo},
\newblock {\em Phys. Rev.} {\bf 127}, 1999 (1962).

\end{thebibliography}

\end{document}